\documentclass[prl, twocolumn,10pt]{revtex4-1}
\usepackage{graphicx}
\usepackage{amsmath}
\usepackage{amsfonts}
\usepackage{dsfont}
\usepackage{amssymb}
\usepackage{color}
\usepackage{comment}
\usepackage[colorlinks,linkcolor=blue,citecolor=blue]{hyperref}

\begin{document}
\title{Reply to `Comment on ``Absence versus Presence of Dissipative Quantum Phase Transition in Josephson Junctions'''}
\author{Kanta Masuki}
\email{masuki@g.ecc.u-tokyo.ac.jp}
\affiliation{Department of Physics, University of Tokyo, 7-3-1 Hongo, Bunkyo-ku, Tokyo 113-0033, Japan}

\author{Hiroyuki Sudo}
\affiliation{Department of Physics, University of Tokyo, 7-3-1 Hongo, Bunkyo-ku, Tokyo 113-0033, Japan}

\author{Masaki Oshikawa}
\affiliation{Institute for Solid State Physics, University of Tokyo, Kashiwa, Chiba 277-8581, Japan}
\affiliation{Kavli Institute for the Physics and Mathematics of the Universe (WPI),
  University of Tokyo, Kashiwa, Chiba 277-8583, Japan}

\author{Yuto Ashida}
\email{ashida@phys.s.u-tokyo.ac.jp}
\affiliation{Department of Physics, University of Tokyo, 7-3-1 Hongo, Bunkyo-ku, Tokyo 113-0033, Japan}
\affiliation{Institute for Physics of Intelligence, University of Tokyo, 7-3-1 Hongo, Bunkyo-ku, Tokyo 113-0033, Japan}

\maketitle

In our Letter~\cite{Masuki_absence_2022}, we determined the ground-state phase diagram of the resistively shunted Josephson junction (RSJ) on the basis of two independent nonperturbative renormalization group (RG) analyses, namely, numerical RG (NRG) and functional RG (FRG). Our main finding is that the insulating phase is strongly suppressed to deep charging regimes due to the dangerously irrelevant term. In Ref.~\cite{Masuki_absence_2022}, we first benchmarked our NRG scheme by making comparisons with the known results in the boundary sine-Gordon model, a well-established theory in condensed matter physics. Then, we applied the scheme to the case of the exact Hamiltonian of RSJ including the capacitance term \(\nu\). We found that \(\nu\) is a dangerously irrelevant term which triggers the nonmonotonic RG flow of the phase mobility, leading to a qualitatively different phase diagram from the commonly believed one  (Fig.~1(a) in Ref.~\cite{Masuki_absence_2022}). Importantly, our independent FRG analysis also validated all the qualitative features found in the NRG analysis.

The Comment~\cite{Sepulcre_comment_2022} by S{\'e}pulcre, Florens, and Snyman (SFS) is mainly two folds. First, they claim that our NRG scheme is uncontrollable by suggesting that the resulting excitation spectrum can be different from what would be expected when approximating the cosine potential as \(-\epsilon_J\cos(\hat\Xi) \approx \epsilon_J(\hat\Xi^2-1)/2\). Second, SFS argue that there seems to be no clear evidence for the reentrant transition in the limit \(\alpha\to +0\). After carefully studying the comments, we conclude that none of them affect the main finding of our Letter, that is, the strong suppression of the insulating phase into deep charging regimes. We emphasize that all the comments are concerned with NRG analysis, and the FRG analyses in the Letter (which unambiguously support the NRG results) remain intact.

\begin{figure}[h]
  \includegraphics[width=0.48\textwidth]{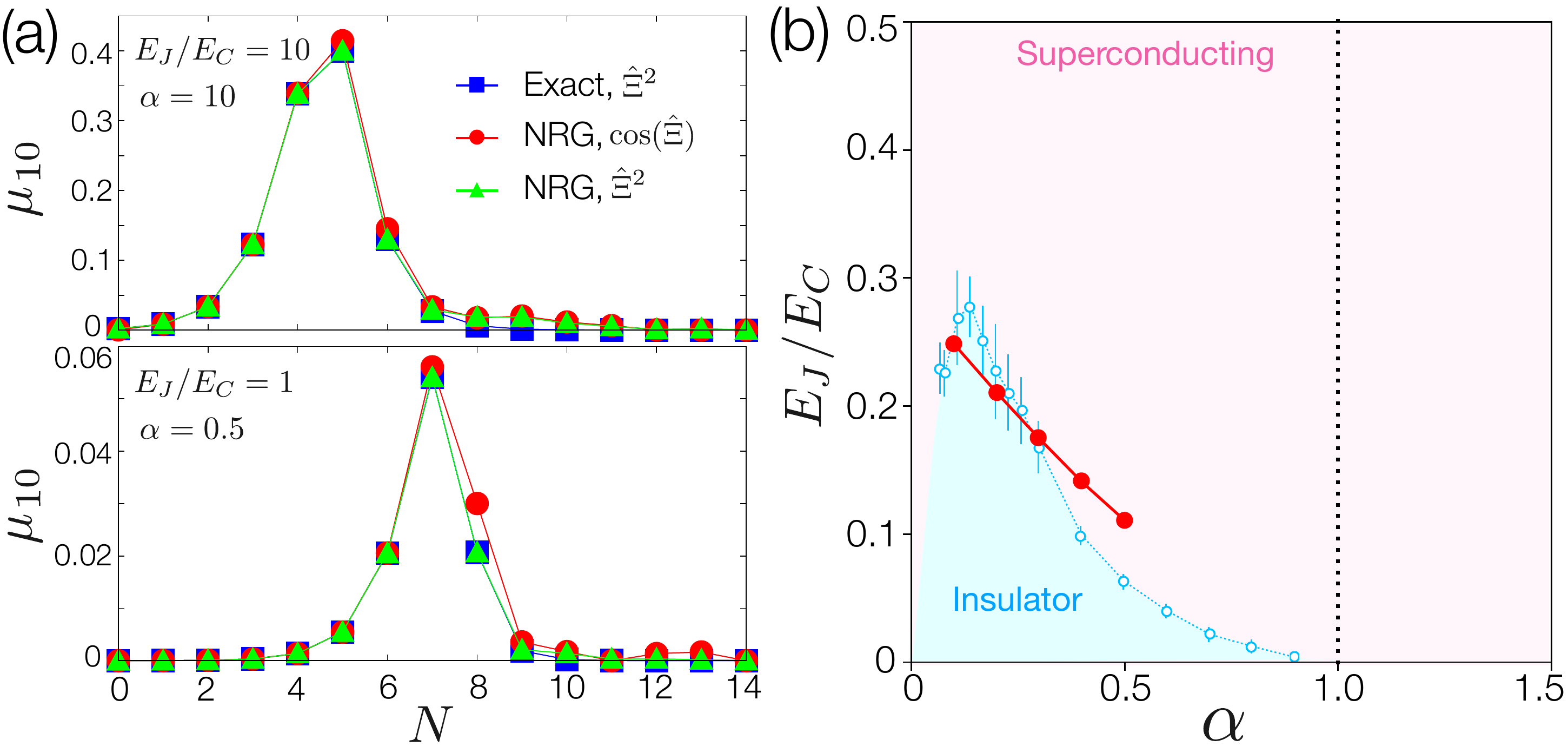}
  \caption{(a) Phase mobility \(\mu_{10}\) plotted against the number of NRG steps \(N\). Parameters are \(E_J/E_C\!=\!10\) and \(\alpha\!=\!10\) in the top panel (the same with Fig.~1 in Ref.~\cite{Sepulcre_comment_2022}) and \(E_J/E_C\!=\!1\) and \(\alpha\!=\!0.5\) in the bottom panel. We take \(E_C\!=\!0.01\) and \(\Lambda\!=\!2.0\) in both panels. (b) Ground-state phase diagram of RSJ reproduced from Ref.~\cite{Yokota_Functional_2022}. The red points (blue circles) are the phase boundaries obtained from the FRG analysis presented in Ref.~\cite{Yokota_Functional_2022} (the NRG~\cite{Masuki_absence_2022}).\label{figure}}
\end{figure}

In the top panel of Fig.~\ref{figure}(a), we plot the phase mobility \(\mu_{10}\) obtained by different methods: NRG for the cosine potential \(-\epsilon_J\cos(\hat\Xi)\), NRG for the quadratic potential \(\epsilon_J\hat \Xi^2/2\), and the exact diagonalization for the the quadratic potential. We here choose the same parameters as in the Comment. All of the results agree well with each other and converge to zero, which is consistent with the fact that the system is in the superconducting phase. This quantitative agreement shows that our NRG analysis can faithfully determine the values of order parameters at the ground state, thus supporting the validity of our procedure for locating the transition point on the basis of calculating the phase mobility. Indeed, the nonmonotonic behavior confirmed in Fig.~\ref{figure}(a) is the hallmark of the dangerously irrelevant behavior discussed in Ref.~\cite{Masuki_absence_2022}.
As a further confirmation, we also plot the phase mobility at \(\alpha = 0.5\) (bottom panel in Fig.~\ref*{figure}(a)), where again all the results agree very well and flow to the superconducting phase, which is precisely consistent with our main finding that the system always remains  superconducting at large \(E_J/E_C\) even for $\alpha<1$.  We found similar agreements in calculations of the phase coherence \(\langle\cos(\varphi)\rangle\).
It is also useful to recall that in Ref.~\cite{Masuki_absence_2022} we benchmarked the NRG analysis by confirming the well-known transition point at $\alpha_c=1$ in the boundary sine-Gordon model with an error less than $\sim 1\%$.

Regarding the second comment, at least in the current implementation of NRG, it is  technically challenging to accurately analyze the regime  \(\alpha\ll 1\), as we have already explained in the Supplementary Materials of Ref.~\cite{Masuki_absence_2022}. This is the reason why error bars in the phase diagram were relatively large and the phase boundary was indicated only by a dashed extrapolated curve in this region (see Fig.~1(a) in Ref.~\cite{Masuki_absence_2022}). In this sense, we agree that there is room for further improving quantitative accuracy of NRG scheme at $\alpha\ll 1$. Nevertheless, we disagree with SFS's argument that there is no reentrant transition. This is simply because the absence of the reentrant transition leads to a counterintuitive result that a phase transition should occur at nonzero \((E_J/E_C)_c\) for \(\alpha = +0\); such behavior is not expected since, when \(\alpha\) is zero, the junction is completely decoupled from the environment and there merely exists a single particle in cosine potential, which should clearly not exhibit any phase transition.

Lastly, we recall that all the qualitative features found in NRG analysis, such as the dangerously irrelevant behavior and the suppression of the insulating phase, precisely agreed with the ones revealed by the independent FRG analysis. In a recent work ~\cite{Yokota_Functional_2022}, this agreement has been now further confirmed at the quantitative level by employing a more advanced FRG analysis than the one originally performed in Ref.~\cite{Masuki_absence_2022}. For the sake of completeness, we show the obtained phase diagram (Fig.~\ref{figure}(b)), which gives a further support for the validity  of our NRG analysis.

\begin{acknowledgments}
  Y.A. acknowledges support from the Japan Society for the Promotion of Science through Grant Nos.~JP19K23424 and JP21K13859.
  M.O. is supported in part by MEXT/JSPS KAKENHI Grant
  Nos.~JP17H06462 and JP19H01808, JST CREST Grant
  No.~JPMJCR19T2.
\end{acknowledgments}

\bibliography{references}

\end{document}